# Time-dependent THz dielectric function of ZnTe under two-photon optical excitation at 800 nm wavelength

Farell Keiser[1,2,*], Wentao Zhang[1], Dominik Johannesmann[1],
Nicolas S. Beermann[1], Yuhao Meng[1], Hassan A. Hafez[1], Savio Fabretti[1],
and Dmitry Turchinovich[1,**]

[1] *Fakultät für Physik, Universität Bielefeld, Universitätsstraße 25, 33615 Bielefeld, Germany*
[2] *Gymnasium Leopoldinum, Hornschestraße 48, 32756 Detmold, Germany*

*Corresponding authors:* *fkeiser@physik.uni-bielefeld.de **dmtu@physik.uni-bielefeld.de

**Abstract:**

ZnTe is arguably the most widely used nonlinear crystal for the generation and detection of THz radiation, used in conjunction with sub-bandgap optical excitation by femtosecond lasers operating near 800 nm. The THz dielectric function of ZnTe is the key parameter defining the efficiency and bandwidth of THz generation and detection. Here, we demonstrate that the THz dielectric function of ZnTe undergoes substantial transient modification at 800 nm sub-bandgap excitation under conditions typical for THz generation. These modifications arise from significant free-carrier generation via two-photon absorption of the 800 nm pump, accompanied by the pump-driven activation of the THz-active phonon modes. Using optical pump–THz probe spectroscopy, we characterized the THz dielectric function of ZnTe under 800 nm excitation as a function of pump fluence and pump-probe delay. Analysis of the experimental data within the Drude–Lorentz model provided the generated free-carrier density and momentum-scattering time, and oscillator strength of the pump-activated THz phonon modes, revealing their transient evolution in dependence on the excitation conditions.

## I. Introduction

Nonlinear crystals are widely employed in laser-based THz generation and detection schemes, such as optical rectification and electro-optic sampling, respectively (see [1,2] and references therein). The efficiency and bandwidth of THz generation and detection are determined by the magnitude of the second-order nonlinear coefficient of the crystal, by the transparency of the crystal at both optical laser and THz frequencies, as well as by the phase-matching – the synchronized co-propagation of the optical laser signal and the generated or detected THz signal within the crystal (for the overview see [1–3] and references therein).

Since its first demonstration [4], zinc telluride (ZnTe) has remained the most widely used nonlinear crystal for THz generation and detection. This popularity stems from its compatibility with femtosecond lasers operating near 800 nm, such as Ti:sapphire and frequency-doubled telecom-band lasers that represent the most prevalent femtosecond systems available today [1]. ZnTe with its bandgap of 2.26 eV [5–9] is nominally transparent at the laser wavelength of 800 nm corresponding to the photon energy of 1.55 eV and has a relatively large nonlinear coefficient $r_{41} = 4.04\ \mathrm{pm\ V^{-1}}$ [10]. Crucially, ZnTe provides for the broadband collinear phase-matching between the 800 nm laser pulses and the generated or detected THz signals, ensuring a straightforward experimental implementation of such a THz generation and detection scheme.

The key physical parameter defining the efficient optical-to-THz phase-matching of the nonlinear crystal is its THz dielectric function $\tilde{\varepsilon}(f) = \tilde{n}(f)^2$, where $\tilde{n} = n + ik$ is the complex refractive index, and $f$ is the THz frequency [3,11]. The THz dielectric function of intrinsic ZnTe was thoroughly investigated in the past in the linear stationary regime [12–15], and was found to be dominated by the transverse optical (TO) phonon mode centered at $f_{TO} =$ 5.32 THz at room temperature [13]. This stationary dielectric function of ZnTe has been traditionally used for the mathematical modelling of the optical rectification and electro-optic sampling processes in ZnTe (see e.g. [3,11,13]).

The studies of THz generation in ZnTe under power-dependent 800 nm excitation [16,17] have, however, shown that stronger optical excitation reduces the THz generation efficiency and narrows the generated THz bandwidth. These detrimental effects have been qualitatively attributed to two-photon absorption (TPA) of the 800 nm light in ZnTe [16–18], which leads to pump pulse depletion and generation of free carriers that subsequently reabsorb the THz radiation generated within the crystal. Nevertheless, the broadband complex THz dielectric function of ZnTe under 800 nm excitation conditions relevant for THz generation has not yet been reported or comprehensively analyzed, to the best of our knowledge.

In this work, we perform an 800 nm pump–THz probe experiment on ZnTe to directly characterize its dielectric function under conditions relevant to THz generation. We show that sub-bandgap excitation at 800 nm not only causes pump depletion and free-carrier generation but also modifies the THz-active phonon spectrum on sub-picosecond and picosecond timescales, thereby affecting the THz dielectric function of ZnTe on a timescale relevant for the THz generation.

This paper is structured as follows: Section II describes the details of the optical pump–THz probe experiment. In Section III we present the data analysis using the Drude-Lorentz model, and extract the free-carrier and phonon contributions to the THz dielectric function of the photoexcited ZnTe. In Section IV, we quantify the TPA in ZnTe at the pump wavelength of 800 nm using the fluence-dependent optical transmission measurements, which directly yields the density of photogenerated free carriers in ZnTe. We further use this information to disentangle the contributions of free carrier density and momentum-scattering time to the THz-measured Drude conductivity of photoexcited ZnTe. Section V is dedicated to the discussion of phonon mode activation and carrier-phonon coupling in ZnTe under 800 nm sub-bandgap excitation. Finally, we present our conclusions in Section VI.

## II. Experimental details

As a sample, we used a nominally intrinsic (110)-oriented ZnTe crystal with lateral dimensions of $10\ \mathrm{mm} \times 10\ \mathrm{mm}$ and a thickness of $d = 0.5$ mm, acquired from Eksma Optics UAB. In order to obtain the THz frequency-dependent dielectric function of our sample, the optical pump–THz probe (OPTP) spectroscopy was performed [1,19–21]. Our experiment was powered by an amplified Ti:sapphire laser operating at a 1 kHz repetition rate, which delivered pulses with an 800 nm central wavelength (see Appendix Fig. A1) and a 103 fs duration (FWHM, $1/e^2$) as determined by an autocorrelation measurement. The 800 nm pump pulses in our experiment had the pump fluence range of $0.02 - 0.98\ \mathrm{mJ\ cm^{-2}}$. The THz probe pulses were generated by optical rectification of 800 nm laser pulses in a (110)-oriented ZnTe crystal, and were detected using the free-space electro-optic sampling in yet another (110)-oriented ZnTe crystal. Both THz emitter and detector crystals had the thickness of 1 mm, providing the useful spectroscopy window of 0.5 – 2.5 THz. The peak field strength in the generated THz probe pulse did not exceed $1\ \mathrm{kV\ cm^{-1}}$, thus ensuring the linear THz spectroscopy regime.

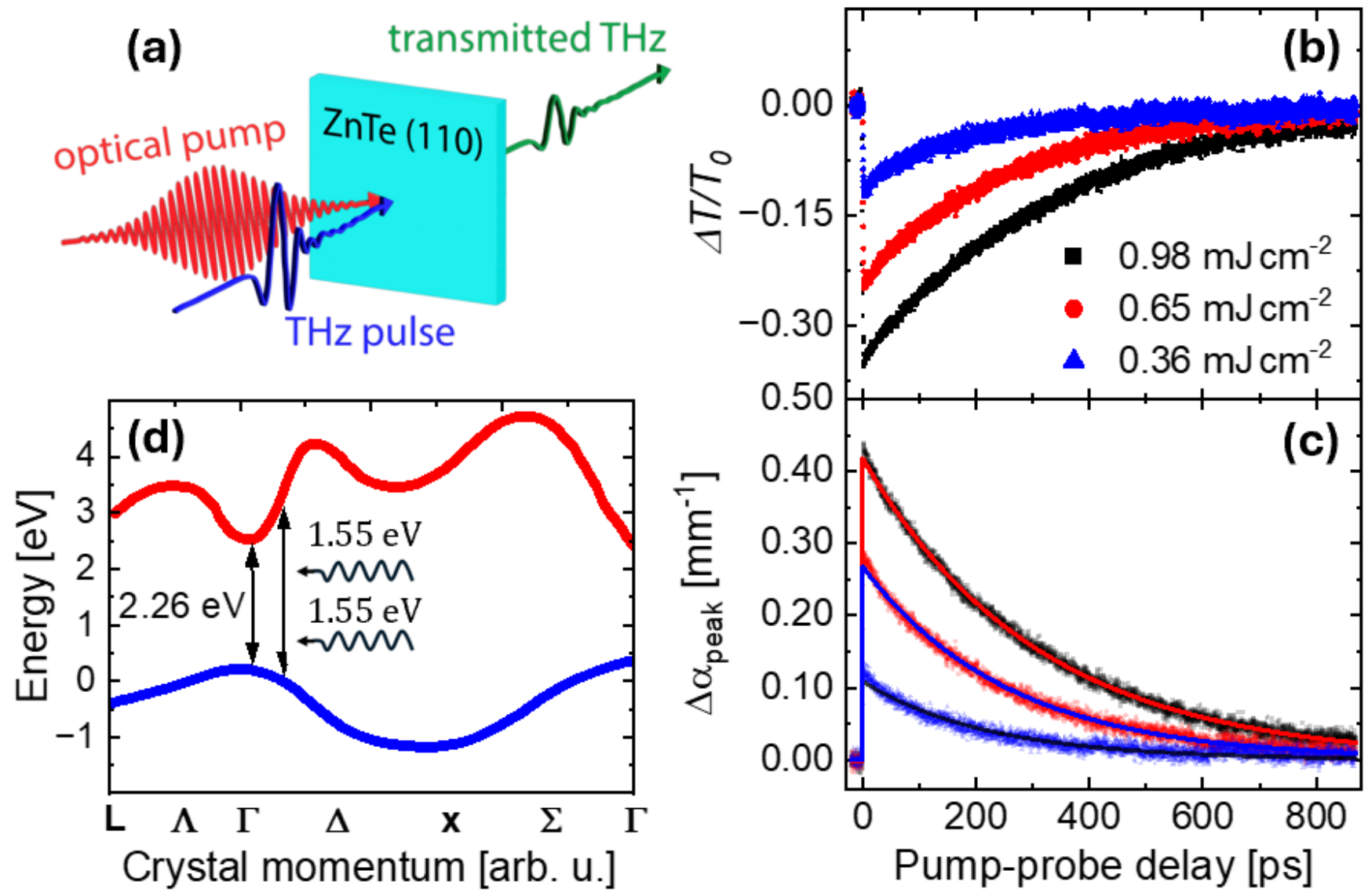

Fig. 1: (a) Schematic of the optical pump–THz probe measurement. (b) The relative change in transmission of the THz peak field through the ZnTe crystal for different pump-probe delays between optical pump and THz probe. (c) Calculated THz frequency-averaged absorption coefficient of 800 nm - excited ZnTe for different pump-probe delays. (d) Illustration of the bandstructure of ZnTe around bandgap, reproduced from [9]. The smaller arrow indicates the band gap energy of 2.26 eV. The larger arrow schematically illustrates the two-photon absorption process at 800 nm pump wavelength corresponding to photon energy of 1.55 eV.

The optical pump–THz probe spectroscopy was performed in a collinear transmission configuration, where both the 800 nm optical pump and the THz probe pulses were incident onto the ZnTe sample at normal incidence, and the transmitted THz probe pulse was detected [1,19–21], as schematically shown in Fig. 1(a). The remainder of the 800 nm pump beam transmitted through the sample was blocked by a black polyethylene foil (acquired from Origin Co., Ltd., Japan), acting as a near-infrared cut-off filter which is opaque to the optical pump but highly transparent to the THz probe pulse. The THz signals in our experiment were detected using a standard lock-in detection scheme, whereas the laser beam used for the THz generation was modulated by an optical chopper. This ensured that only the THz field originating from the THz emitter crystal and intended for the THz spectroscopy, but not the THz emission from the optically excited ZnTe sample crystal itself, was detected in our experiment. All our measurements were performed at room temperature. To prevent the THz absorption by atmospheric water, our spectrometer was purged with dry nitrogen. In order to exclude any possible THz diffraction effects in OPTP, the optical pump beam spot on the sample was kept at the size of about three times larger than the THz probe beam spot. To ensure the high stability of our experiment, the optical path of the 800 nm laser was continuously monitored and controlled by an active beam stabilizer.

The THz data analysis was performed in the following fashion. Given that ZnTe at 800 nm pump wavelength is nominally transparent, and the only possible mechanism of free carrier generation is TPA, the sample was assumed to be uniformly excited within its entire thickness. Consequently, classical THz time-domain spectroscopy (THz-TDS) equations [1,19–21] were used for the calculation of the THz dielectric function from the measured THz transmission. These calculations are based on the verified assumption that the real part of the refractive index largely exceeds the imaginary part (see Fig. 2(a)). Consequently, the amplitude of the complex transmission function depends solely on the real refractive index, whereas its phase is governed by the full complex refractive index. We further note here that the optical pump creates a penetration-depth-dependent non-uniform excitation profile within the ZnTe crystal. However, all physical quantities extracted from our measurements are averaged over the entire crystal thickness as effective characteristics, see Section A1 in the appendix.

As a THz-TDS reference, the probe THz pulse propagated through the spectrometer without the sample in place was recorded. In order to exclude the “ageing” of the probe THz pulses in the OPTP measurement, the time-delay between the 800 nm optical pump and THz detection optical arms of the spectrometer was kept constant, and the delay between these two and the THz generation arm was varying. Such an approach ensures that all the data points within the detected THz probe waveform correspond to one and the same pump-probe delay [1,19–21].

The sample and reference measurements were periodically alternated by moving the sample in and out of the THz beam using an automated translation stage, thus minimizing any possible long-term laser drift effects on our experiment.

**III. Transient THz absorption and dielectric function**

To investigate the THz absorption dynamics in ZnTe subsequent to 800 nm excitation, we first measure the electric field at the peak of the transmitted THz probe pulse as a function of the pump-probe delay [22], at the pump fluence varying in the range $0.36-0.98$ mJ cm$^{-2}$. Fig. 1(b) displays the relative change in THz transmission – the pump-induced change in the field transmission $\Delta T$ normalized to the field transmission measured before excitation $T_0$. We observe a pronounced reduction in THz transmission at the point of temporal overlap of the 800 nm pump and THz probe pulses, followed by a relaxation on a timescale of 100s of ps. This reduction of the THz transmission in an 800 nm-excited ZnTe is the signature of a significant free carrier generation in ZnTe by TPA, leading to increased crystal conductivity and, consequently, to enhanced THz absorption [16–18]. In the following section, we will quantify this effect using the Drude conductivity model.

The change in the frequency-averaged THz power absorption coefficient $\Delta\alpha_{\mathrm{peak}}$ is determined from the field transmission data in Fig. 1(b) using Beer's law $T = e^{-\alpha d/2}$ and is presented in Fig. 1(c). To extract the lifetime $\tau_{rel}$ of the free-carrier conductivity, we fit the measured $\Delta\alpha_{\mathrm{peak}}$ with an exponential decay function $\Delta\alpha_{\mathrm{peak}}(t) \propto \exp(-t/\tau_{rel})$, where *t* is time. At pump fluences of $F_p = 0.36$ mJ cm$^{-2}$, $F_p = 0.65$ mJ cm$^{-2}$ and $F_p = 0.98$ mJ cm$^{-2}$, the corresponding photoconductivity lifetimes were found to be $\tau_{rel} = 224$ ps, $\tau_{rel} = 258$ ps, and $\tau_{rel} = 308$ ps, respectively. We assume, that the photoexcitation predominantly populates the Γ-valley of the conduction band of ZnTe (see Fig. 1(d)). In the Γ-valley, the electron effective mass of $m_e^* = 0.116\, m_0$ is significantly smaller than that of the heavy hole $m_h^* = 0.69\, m_0$ [23,24], leading to the negligibly small hole mobility as compared to electrons [25,26]. Therefore, the measured photoconductivity in ZnTe can be assigned to electrons only. Below in Section IV, we will show that the momentum scattering time in ZnTe is strongly dependent on the free carrier density: it decreases monotonously and nonlinearly with the electron density. Consequently, the measured conductivity dynamics in Fig. 1(c) does not represent exclusively the free carrier recombination time, but rather a convoluted effect of the dynamics of the electron density and density-dependent electron mobility. Consequently, while the instantaneous electron mobility is governed by ultrafast scattering events, it is the carrier recombination time that causes the observed long-lived relaxation dynamics.

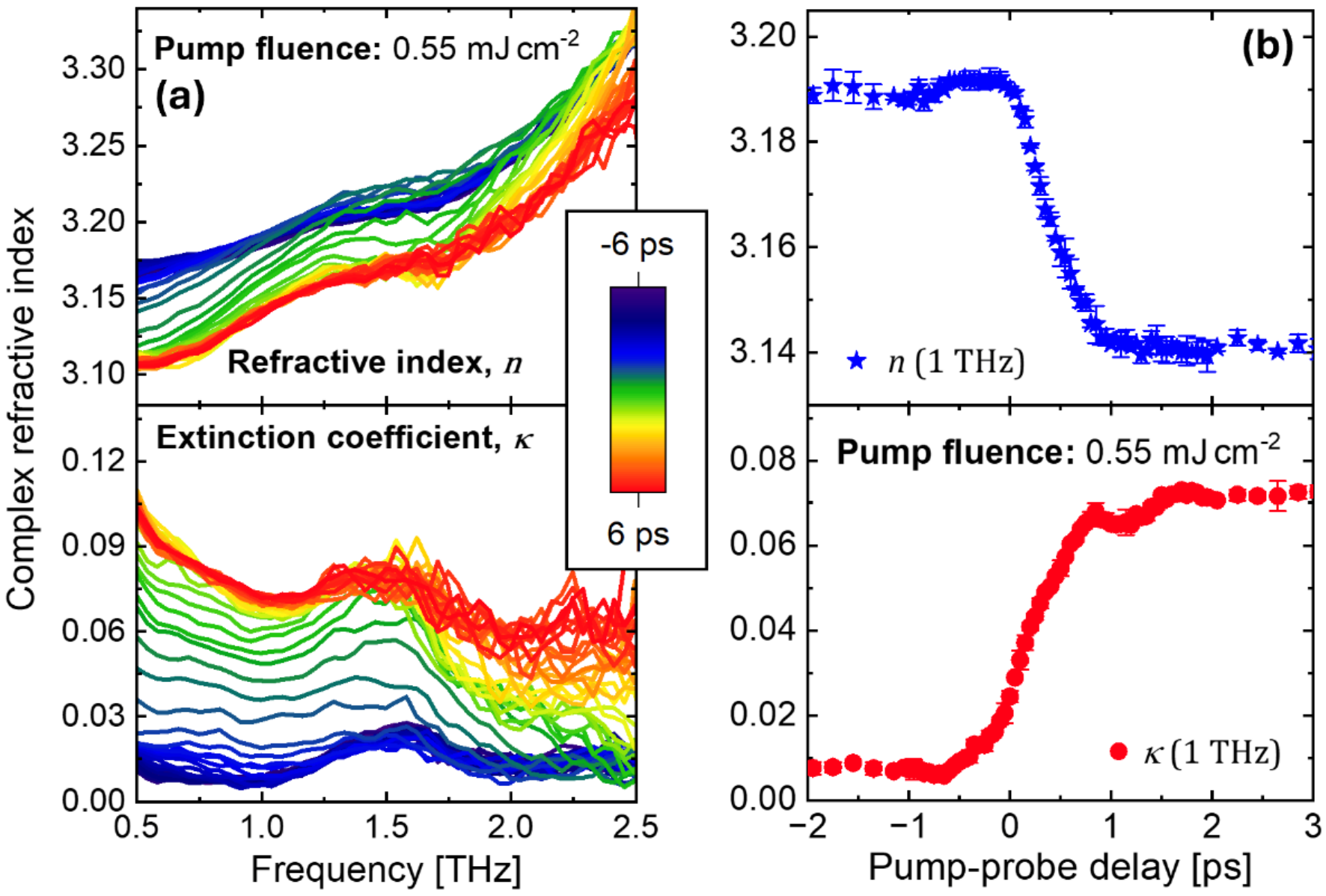

Fig. 2: (a) The frequency-dependent complex refractive index at selected pump-probe delays from -6 to 6 ps. The refractive index is displayed in the top, the extinction coefficient in the bottom part of the graph. (b) The dynamics of the complex refractive index at a frequency of 1 THz and a pump fluence of 0.55 mJ cm$^{-2}$.

The OPTP measurement allows us to extract the complex THz frequency-dependent refractive index of ZnTe following the photoexcitation, $\tilde{n}(f) = n(f) + ik(f)$ [1,19–21]. In Fig. 2(a), we show the temporal evolution of the measured spectra of the real part of the refractive index $n(f)$, in the following referred to as the refractive index, and of the extinction coefficient $\kappa(f) = \frac{\alpha(f)c}{4\pi f}$ in the range of pump-probe delay times from $-6$ ps to 6 ps, measured at the 800 nm excitation fluence $F_p = 0.55$ mJ cm$^{-2}$. Here $\alpha(f)$ is the THz frequency-dependent power absorption coefficient, and $c$ is the speed of light in vacuum.

Before the optical excitation, we observe significant dispersion in the THz refractive index spectrum of ZnTe, in full agreement with previous linear characterization studies [12–15]. This dispersion primarily arises from the TO phonon at $f_{TO} = 5.32$ THz, which marks the beginning of the reststrahlen band in ZnTe [13,15]. In addition to the dominating refractive index dispersion associated with the TO phonon, we also clearly resolve a resonance at $f_{L1} = 1.55$ THz. Earlier studies have shown two low-frequency phonon modes at $f_{L1} = 1.55$ THz and $f_{L2} = 3.69$ THz [13] in ZnTe and identified them as combination phonon modes [13,15]. The TO phonon mode at $f_{TO} = 5.32$ THz and the phonon mode at $f_{L2} = 3.69$ THz lie outside the spectral window of our experiment.

As clearly seen in Fig. 2(a), the refractive index $n(f)$ of ZnTe generally decreases, whereas the extinction coefficient $\kappa(f)$ generally increases as a result of the 800 nm excitation. At lower THz frequencies $\kappa(f)$ diverges, indicating a non-zero dc conductivity due to photogenerated free carriers which will be analyzed in the following section. To emphasize the pronounced temporal dynamics of $n(f)$ and $\kappa(f)$, Fig. 2(b) shows their values at a representative frequency of $f = 1$ THz as a function of pump-probe delay. Both quantities undergo substantial changes on the timescale of ~1.5 ps: the refractive index decreases by ~5 % whereas the extinction coefficient increases by a factor of $\sim 8$ This timescale of ~1.5 ps manifests the establishment of maximum conductivity within the photo-excited free carrier population via electron-electron and electron-phonon correlations, see for example [22,27]. When THz pulses are generated in ZnTe via optical rectification of 800 nm pulses, the duration of the resulting THz pulse is typically on the order of such timescales. Therefore, the medium's refractive index and absorption coefficient are not stationary during the actual THz generation and propagation process. This dynamically alters the phase-matching condition and reabsorption, directly affecting the emitted THz waveform.

### A. Drude-Lorentz description of the measured dielectric function

To quantitatively describe the time- and pump-fluence-dependent dielectric function of ZnTe $\tilde{\varepsilon}(\omega) = \tilde{n}(\omega)^2$ , we fit the measured data using the multi-term Drude-Lorentz model [28,29].

$$\tilde{\varepsilon}(\omega) = \varepsilon_\infty - \frac{\sigma_{dc}/\varepsilon_0}{\tau_e\omega^2 + i\omega} + \Omega_{L1}^2 \frac{\omega_{L1}^2}{\omega_{L1}^2 - \omega^2 - i\omega\gamma_{L1}}$$

$$+\Omega_{L2}^2 \frac{\omega_{L2}^2}{\omega_{L2}^2 - \omega^2 - i\omega\gamma_{L2}} + \Omega_{TO}^2 \frac{\omega_{TO}^2}{\omega_{TO}^2 - \omega^2 - i\omega\gamma_{TO}}. \qquad (1)$$

Here $\varepsilon_\infty$ is the dielectric constant at high frequency; the second term in Eq. 1 represents the Drude function accounting for the contribution of the free carriers to the dielectric response [29], where $\tau_e$ is the carrier momentum scattering time and $\sigma_{dc} = \frac{e^2\tau_e N}{m^*}$ is the dc conductivity, where $e$ is the electron charge, $N$ is the free electron density, and $m^*$ is the electron effective mass. The third, fourth and fifth Lorentzian terms in Eq. 1 describe the phonon modes at $f_{L1} = 1.55$ THz, $f_{L2} = 3.69$ THz and $f_{TO} = 5.32$ THz, with $\omega_{L1,L2,TO} = 2\pi\, f_{L1,L2,TO}$ being the center angular frequency, $\Omega_{L1,L2,TO}$ the oscillator strength, and $\gamma_{L1,L2,TO}$, the linewidth of the corresponding phonon mode.

Because of the high density of states associated with the fundamental TO phonon mode and the (110) crystal orientation used in our experiments, it is reasonable to assume that this mode will not undergo significant softening or hardening as a result of photoexcitation. Consequently, in our fitting we use the fixed values of its center frequency $f_{TO}$= 5.32 THz, linewidth $\frac{\gamma_{TO}}{2\pi} = 0.025$ THz, and oscillator strength $\Omega_{TO}^2$ = 2.58, as previously reported under linear static conditions in Ref. [13].

The phonon mode at $f_{L2} = 3.69$ THz contributes only weakly to the overall dielectric function of ZnTe [13,15]. However, it is known that this mode can be activated by optical excitation [30–32]. Since our experiment does not frequency-resolve this mode completely, in our fitting we fix its center frequency to the reported value $f_{L2} = 3.69$ THz [13].

Generally, our THz spectroscopy window falls into the regime of $\tau_e \omega \ll 1$. In this limit, the dispersion of the experimentally measured conductivity spectra is too weak to reliably extract the scattering time $\tau_e$ via unconstrained fitting. To ensure high numerical stability of our fitting, we were therefore prompted to fix it as a constant parameter within our fitting routine. Consequently, we fixed $\tau_e$ = 35.9 fs, which is the value of $\tau_e$ at low carrier densities–determined by the combination of THz spectroscopy and optical transmittance measurements, as explained below.

We fit the measured frequency-dependent THz dielectric function of ZnTe using Eq. 1 to extract the relevant material parameters as functions of THz frequency, 800 nm pump fluence, and pump-probe delay. A selection of our fitting results is displayed in Fig. 3. In total, our fitting model comprises 12 parameters, 6 of which were fixed to literature values as described above. A rigorous covariance analysis reveals that all but two of these free parameters are completely independent, whereas the high-frequency dielectric constant $\varepsilon_\infty$ and the oscillator strength $\Omega_{L2}$ of the L2 mode exhibit a strong mathematical anti-correlation. As discussed in detail in Section A2 in the Appendix (see Fig. A2(b)), this degeneracy arises from the limited spectral probing window, causing these two parameters to act as a coupled phenomenological offset to the real part of the dielectric function. While their isolated absolute values cannot be uniquely disentangled without further constraints, their combined effective response perfectly captures the broadband background dynamics. Crucially, this localized cross-talk does not compromise the statistical independence of the remaining free parameters, ensuring that the dynamically relevant features - such as the Drude response and the L1 mode characteristics - are robustly and unambiguously determined.

Fig. 3(a) shows the measured real and imaginary parts of the refractive index spectrum in ZnTe prior to and after the 800 nm excitation with the fluence of $0.55\ \mathrm{mJ\ cm^{-2}}$, at pump probe delays of -2 ps and 1.3 ps, respectively, along with the corresponding fits using Eq. 1. In the upper panel of Fig. 3(a), it can be seen that the optical excitation leads to a significant steepening in the refractive index dispersion of ZnTe. While these drastic dynamics could, in principle, originate from the modification of either the 3.69 THz or the 5.32 THz phonon mode, previous studies strongly point towards the former. While this mode was firstly attributed to originate from a Te segregation [31] or to be a combination mode [30], it has been conclusively attributed to be a two-phonon-bound-state (TPBS) [32]. This identification as a TPBS (Ref. [32]) explains the selective activation observed in our experiment. For the (110)-oriented ZnTe crystal used in our work—which exactly matches the crystal orientation investigated in Ref. [32]—the fundamental TO-phonon at 5.32 THz cannot be efficiently excited via impulsive stimulated Raman scattering due to strict symmetry-based selection rules. In contrast, the TPBS at 3.69 THz is formed by two acoustic phonons with opposite momenta ($q$ and $-q$), thus conserving the total momentum to zero. As demonstrated in Ref. [32], this TPBS exhibits a full Γ1 symmetry, making it polarization-independent and allowing for highly efficient optical excitation by the pump pulse.

Accordingly, optical-pump optical-probe experiments have demonstrated the pump-induced activation of the 3.69 THz mode [31,32], exhibiting relaxation dynamics that closely mirror those observed in our study (see Appendix Fig. A3(b)). Thus, we assume that the fundamental TO phonon mode is inherently less susceptible to such transient changes than the L2 combination mode, owing to the significantly higher density of states of the TO phonon and the (110) crystal orientation. Therefore, in our data analysis, we attribute these dynamics to the transient modification of the Lorentzian parameters of the $f_{L2} = 3.69$ THz combination phonon mode. Following excitation, its oscillator strength initially increases and subsequently exhibits a slight decrease (see Appendix Fig. A3(b)). Conversely, the linewidth of the L2 resonance initially narrows before broadening at later pump-probe delays (see Appendix Fig. A3(c)).

The lower panel of Fig. 3(a) shows the measured and fitted spectra of the extinction coefficient $\kappa(f)$ of ZnTe prior to and after the photoexcitation. The photoexcitation leads to a significant increase of the extinction coefficient primarily at lower THz frequencies. Whereas the dynamics of the refractive index spectrum (upper panel of Fig. 3(a)) are dominated by the $f_{L2} = 3.69$ THz phonon dynamics, the extinction coefficient spectrum (lower panel of Fig. 3(a)) is largely influenced by the free-carrier generation, as will be discussed below.

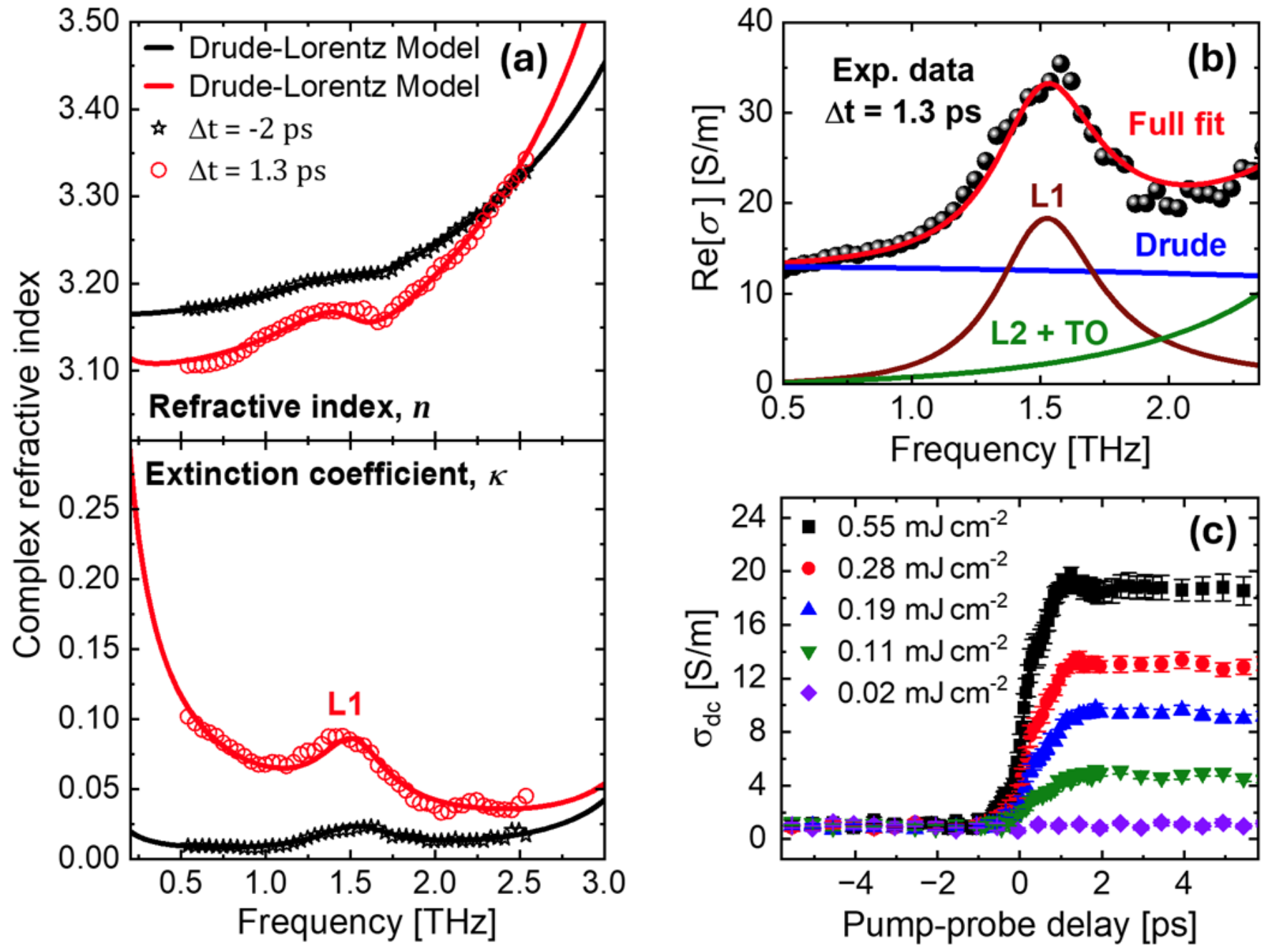

Fig. 3: (a) The measured complex refractive index 2 ps before and 1.3 ps after the excitation, at the pump fluence of 0.55 mJ cm$^{-2}$. The refractive index is depicted in the top, the extinction coefficient in the bottom part of the graph. The experimental data is displayed as symbols and solid lines represent the fitting results. (b) Partial fits in the real part of the conductivity. The black symbols represent the measured data $\Delta t = 1.3$ ps after the excitation with the pump fluence of 0.55 mJ cm$^{-2}$, the red solid line represents the total fit, the blue line the Drude contribution, the reddish line the contribution by resonance L1 and the green line the background conductivity by resonance L2 and the TO-phonon. (c) Dynamics of the dc conductivity for different pump fluences extracted from the fitting model in equation (1).

### B. Carrier dynamics

To provide further insight into the observed dynamics, Fig. 3(b) displays the measured real part of the conductivity spectrum of photoexcited ZnTe at the pump-probe delay of 1.3 ps, given by $\mathrm{Re}[\sigma] = -\omega\varepsilon_0\varepsilon_2$, along with the partial fits corresponding to different terms in Eq. 1. Here $\varepsilon_0$ is vacuum permittivity and $\varepsilon_2$ is the imaginary part of the dielectric function given by Eq. 1. The symbols show the measured data, whereas the solid lines show the isolated contributions from the Drude term, and Lorentz terms describing the L1-phonon resonance, and the combined contributions of the L2 and TO resonances, along with the full fit. To provide a broader visualization of the dielectric function and explicitly illustrate how the low-frequency tails of the higher-frequency modes extend into our measurement window, an extended version of this figure up to 5 THz is provided in Fig. A4 in the Appendix.

This fitting model Eq. 1 is individually applied to all our measured spectra in order to extract the time-dependent evolution of the corresponding parameters defining the free-carrier conductivity and the dynamics of L1 and L2 combination phonon modes. As mentioned previously, the parameters related to the fundamental TO phonon mode remain fixed throughout our fitting routine.

The free-carrier contribution to the measured dielectric function is captured by the Drude term in Eq. 1. The dc conductivity $\sigma_{dc}$, corresponding to the free-carrier conductivity, is obtained directly from the measured data, by extrapolating the real part of the conductivity to zero frequency. Fig. 3(c) shows the established values of $\sigma_{dc}$ of ZnTe excited at 800 nm, with the optical pumping fluence ranging from $0.02$ to $0.55\ \mathrm{mJ\ cm^{-2}}$. The overall decay of the free-carrier conductivity, which is the main contribution to the overall THz absorption of photoexcited ZnTe, occurs on the timescale of many 100s of picoseconds, as shown in Fig. 1(c).

Interestingly, even prior to the optical excitation, a small dc conductivity of $\sigma_{dc} = 1.03 \pm 0.14 \text{ S m}^{-1}$ is observed, indicating the slight background doping of our sample crystal. Considering a carrier momentum scattering time of $\tau_e = 35.9 \pm 3.2$ fs (as determined below in Section IV) and an effective electron mass of $m^* = 0.116\, m_0$ [23], this dc background conductivity corresponds to a doping density of $N = (1.2 \pm 0.2) \times 10^{14} \text{ cm}^{-3}$. Importantly, we have verified that there is no measurable difference between the steady-state spectra (recorded with the optical pump beam completely blocked) and the transient spectra at negative pump-probe delays. Furthermore, this initial dc conductivity is entirely independent of the applied pump fluence. This confirms that the crystal fully relaxes between consecutive laser pulses, ensuring that the measured conductivity is genuinely the result of intrinsic background doping and not an artifact of cumulative carrier accumulation or persistent heating effects.

As discussed above, the carrier momentum scattering time $\tau_e$ could in principle be directly obtained from the dispersion of the Drude conductivity spectrum. However, this dispersion in our spectral window of 0.5 – 2.5 THz is relatively weak (see Fig. 3(b)), indicating the values of $\tau_e$ of the order of ~20 fs with significant error bar [33]. Nevertheless, the significantly more precise determination of carrier momentum scattering time $\tau_e$ becomes possible based on the knowledge of the parameters $\sigma_{dc}$, $N$ and $m^*$, as will be shown below.

**Section IV. TPA of optical pump, free-carrier density and momentum scattering time**

We determined the pump-induced free-carrier density by first measuring the fluence-dependent transmission of the 800 nm beam to calculate its absorption in the ZnTe sample. As shown in Fig. 4(a), the 800 nm beam transmittance through our 1 mm-thick ZnTe crystal drastically reduces from $T \approx 0.46$ down to $T \approx 0.17$ as the 800 nm pump fluence increases from $F_P = 3.48\ \mu\text{J cm}^{-2}$ to $F_P = 1.39 \text{ mJ cm}^{-2}$. Such a strongly nonlinear optical transmission is a signature of TPA in the sample, with the corresponding intensity of the beam transmitted into the sample described by [34,35]

$$I(z, I_0) = \frac{I_0}{1 + \beta z I_0} \qquad (2).$$

Here $\beta$ represents the nonlinear optical absorption coefficient, $z$ is the crystal coordinate and $I_0$ is the intensity incident onto the front face of the crystal located at the longitudinal coordinate $z = 0$. Accounting for reflection losses at the crystal interfaces, we obtain the optical transmittance as a function of incident intensity:

$$T_{opt}(I_0) = \frac{T_{opt,01}^2}{1 + \beta d\, T_{01} I_0} \qquad (3).$$

Here $d$ is the crystal thickness and $T_{opt,01}$ is the optical transmittance of an air-ZnTe interface, assumed to be intensity-independent. By applying Eq. 3 to fit the experimental data, as shown in Fig. 4(a), we extract $T_{opt,01} = 0.678\ \pm 0.002$ and $\beta = 4.30\ \pm 0.09$ cm GW$^{-1}$, whereby the intensity-fluence relation $I_0 = \sqrt{\frac{2}{\pi}} \frac{F_P}{\tau_P}$ was used for a Gaussian pulse with the duration of $\tau_P = 103$ fs (FWHM, $1/e^2$). This value of nonlinear optical absorption coefficient determined by us at the pump wavelength of 800 nm is in reasonable agreement with previously reported values of $\beta = 4.2$ cm GW$^{-1}$ for ZnTe under a slightly different pump wavelength of 1064 nm [36].

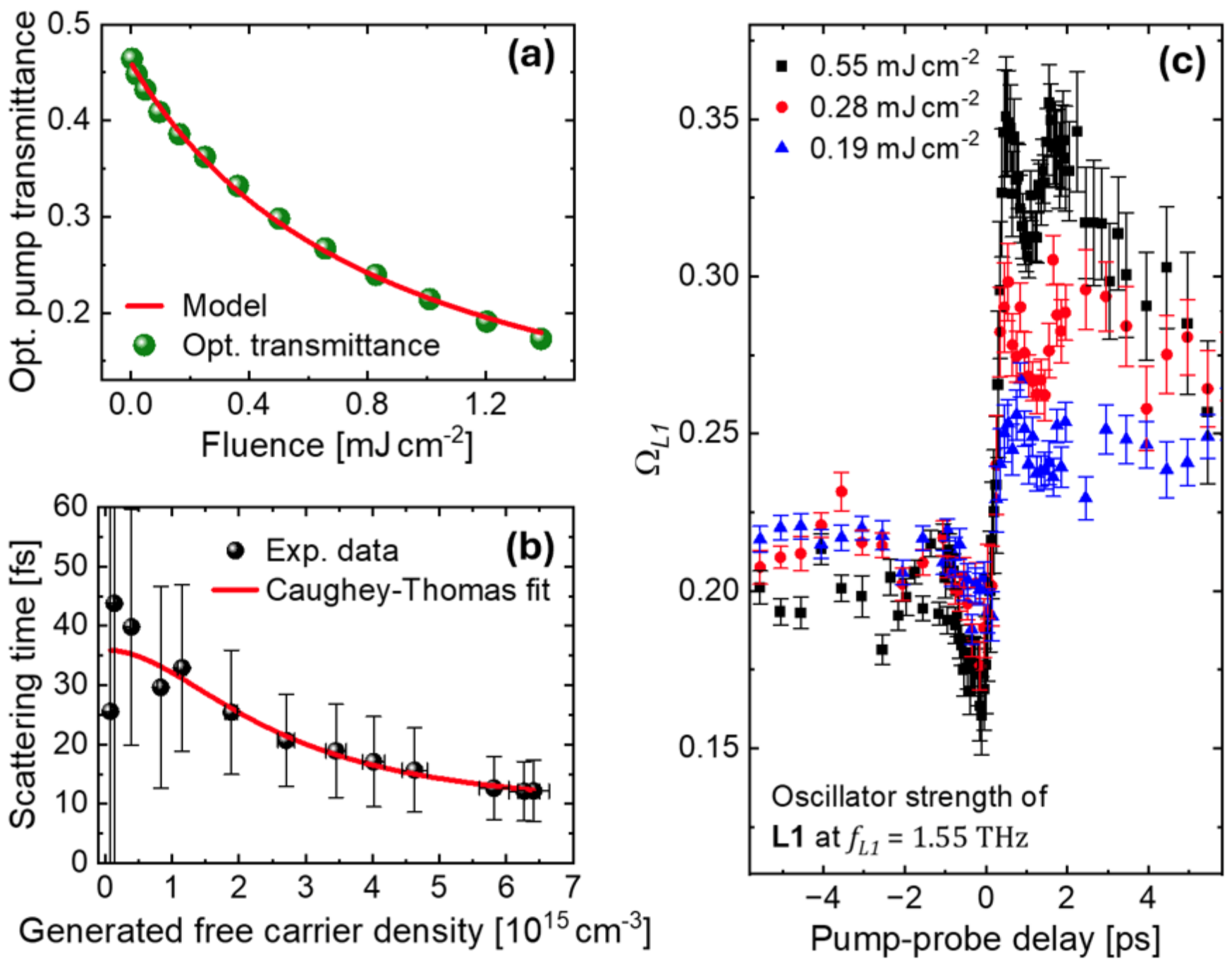

Fig. 4: (a) Optical transmittance of the pump beam measured for different fluences. The solid red line corresponds to the fit with Eq. 3 (b) Obtained scattering time for different generated carrier densities, whereas the solid red line corresponds to the fitting with the Caughey-Thomas model. (c) The oscillator-strength of resonance L1 at 1.55 THz for different pump fluences. The error bars display the standard deviations from the fitting.

Now, from the transmitted intensity, the number of absorbed photons $n_{abs}$ can be calculated, which is directly related to the generated carrier density $N$ via TPA:

$$n_{abs} = A \times \frac{I - T_{01}^2 I_0}{f_{rep} \hbar\omega} = 2\, N\, V \qquad (4)$$

$I$ and $I_0$ correspond to the transmitted and input intensity, respectively, $A$ represents the illuminated area of the sample, $V$ is its illuminated volume, $\hbar\omega$ = 1.55 eV is the photon energy at 800 nm wavelength, and $f_{rep}$ = 1 kHz is the repetition rate of our laser system. Importantly, the generated carrier density describes a value averaged over the entire crystal thickness, which directly corresponds to the similarly thickness-averaged dc conductivity measured by the THz-probe. The factor of two in Eq. 4 arises from the fact that two absorbed photons are required to generate one electron-hole pair. We remind here that we operated under the assumption that only the electrons significantly contribute to the measured conductivity, due to their much higher mobility as compared to the holes in ZnTe. As a result, the photogenerated carrier density *N* for each used 800 nm pump fluence can be readily obtained. Now, from the time- and pump fluence-dependent dc conductivity $\sigma_{dc} = \sigma_0 + \frac{e^2 N \tau_e}{m^*}$, which is already established and shown in Fig. 3(c), we can readily determine the carrier scattering time $\tau_e$ for various photogenerated carrier densities *N*. Here $\sigma_0 = 1.03 \pm 0.14\ \mathrm{S\ m^{-1}}$ denotes the background dc conductivity arising from the slight doping of our ZnTe sample and established earlier. Since the $\sigma_{dc}$ exhibits negligible relaxation within a few-picosecond timescale, we vary the incident pump fluence and extract the corresponding $\sigma_{dc}$ at a pump-probe delay of 2 ps. We further assume a constant effective mass $m^* = 0.116\ m_0$ [23] of free electrons at all times in our experiment. To incorporate the possible contribution of the much heavier holes into our calculations, we further assume an error $\Delta m^* = m_h$ in the Gaussian error propagation for the error bars illustrated in Fig. 4(b). As depicted in Fig. 4(b), a carrier scattering time varies in the range $\tau_e$ = 12.2 – 43.4 fs in the free carrier density range $N = (0.44 - 6.41) \times 10^{15}\ \mathrm{cm^{-3}}$, with shorter carrier scattering time corresponding to higher carrier density, a well-known phenomenon in ZnTe and other semiconductors [25,26,37,38]. This effect results from the increased phase-space filling at higher electron densities leading to stronger electron scattering, and can be empirically described with a Caughey-Thomas formula [39] or one of its derivatives [39]:

$$\tau = \frac{\tau_2 - \tau_1}{1 + (N/N_0)^{\vartheta}} + \tau_1 \qquad (5)$$

The solid line in Fig. 4(b) depicts the Caughey-Thomas fit to our data, yielding the following parameters: $\tau_1 = 35.9 \pm 3.2$ fs, $\tau_2 = 8.3 \pm 11.5$ fs, $N_0 = (2.57 \pm 1.39) \times 10^{15}\mathrm{cm^{-3}}$, and $\vartheta = 1.93 \pm 1.51$, where $\tau_1$ and $\tau_2$ represent the carrier scattering times at low and high electron density, $N_0$ is the saturation carrier density, and $\vartheta$ is an effective Caughey-Thomas parameter [39]. The carrier scattering time established here corresponds to the mobilities in ZnTe of $\mu_e = \frac{e\tau_e}{m^*}$ = 545-126 $\text{cm}^2\ \text{V}^{-1}\text{s}^{-1}$, for the carrier density range $(0.44 - 6.41) \times 10^{15}\ \text{cm}^{-3}$. While literature reports similar absolute mobilities [25,26], those values correspond to intrinsic, steady-state doping electrons at much higher concentrations on the order of $N = (0.05 - 3) \times 10^{18}\ \text{cm}^{-3}$. In contrast, our experiment captures the dynamics of photo-excited electrons roughly 2 ps after optical excitation. This early onset of strong scattering at relatively low densities indicates that the mobility of these transient carriers is heavily suppressed by robust non-equilibrium mechanisms - such as strong intrinsic electron-phonon coupling which happens due to the highly polar lattice of the II-VI semiconductor, which is also reflected in a large Fröhlich coupling constant [5].

**Section V. Phonon mode activation and carrier–phonon coupling**

From the fitting model in Eq. 1, we further extract the oscillator strength $\Omega_{L1}$ of the combination mode phonon at $f_{L1} = 1.55$ THz, shown in Fig. 4(c). Subsequent to optical excitation, a pronounced increase in the oscillator strength of this phonon mode is observed, while its linewidth remains largely unchanged within the limits of the standard deviations: $\gamma/2\pi = 0.59 \pm 0.08$ THz, see Fig. A3(a) in the Appendix. This observed increase in the oscillator strength of the $f_{L1} = 1.55$ THz phonon mode is a signature of the optical phonon activation. Such a phonon mode activation was observed previously in optical-pump optical-probe measurements and was explained by the impulsive stimulated Raman scattering mechanism [30–32]. However, phonons excited via impulsive stimulated Raman scattering typically exhibit lifetimes of only a few picoseconds due to relatively strong damping [30–32,40,41]. Indeed, considering the extracted linewidth of this $f_{L1} = 1.55$ THz mode of $\frac{\gamma}{2\pi} = 0.59 \pm 0.08$ THz, a lifetime of about 1.7 ps would be expected. Despite that, no such relaxation behavior is observed in our results, and the increased oscillator strength persists over many picoseconds (see Fig. 4(c)).

Since the direct Raman activation of this phonon mode is unlikely due to its relatively long lifetime, largely exceeding its inverse linewidth, as discussed above, we instead attribute this phonon mode activation to an enhanced effective transition dipole moment mediated by the transient free-carrier gas. Due to the intrinsic electron-phonon coupling in ZnTe [5], the transient photo-generated free carriers act as an effective antenna for the incoming THz probe field. When the THz probe pulse traverses the crystal, it drives these free carriers, an effect that is macroscopically observed as broadband Drude absorption. However, when the driving THz frequency exactly matches that of the resonant combination mode—specifically at 1.55 THz—the lattice and the driven carriers co-oscillate. In this resonant state, a portion of the absorbed THz energy is efficiently transferred from the driven electrons into the coupled phonon mode. This interaction increases the effective transition dipole moment of the combination mode. Macroscopically, we measure this as an enhanced oscillator strength (and thus enhanced THz absorption) of the 1.55 THz resonance. Additionally, this combination mode could also be influenced by an enhanced vibrational nonlinearity of the crystal, driven by this electron-phonon coupling [42].

It is important to note that this carrier-mediated effect is not limited to the 1.55 THz resonance. We also observe a persistent steepening of the refractive index dispersion at higher frequencies, which originates from the phonon modes at 3.69 THz and 5.32 THz. Just like the L1 oscillator strength, this steepened dispersion persists without significant relaxation over the pump-probe delay of about 6 ps. This implies a simultaneous increase in the oscillator strengths of these higher-frequency modes, further corroborating the presence of a persistent, broadband electron-phonon coupling.

Consequently, we conclude that optical pumping activates the 1.55 THz phonon mode, along with additional modes outside our spectral window, by a mechanism other than the short-lived impulsive stimulated Raman scattering. Instead, this mode activation, characterized by an increase in both the effective polarizability and the transition dipole moment, is sustained by the driven non-equilibrium electron gas.

Finally, it should be noted that for pump-probe delays between 0 and 2 ps, the extracted oscillator strength exhibits observable oscillations, as shown in Fig. 4(c). These transient fluctuations could plausibly originate from instantaneous nonlinear optical effects—such as the Pockels and Kerr effects—driven by the intense optical pump pulse. These cause a rapid modulation of the refractive index that perturbs the fitting routine. However, this effect is strictly confined to the temporal overlap and does not influence the extracted phonon parameters, or the sustained electron-phonon dynamics observed at larger positive pump-probe delays.

**Section VI. Conclusion**

We have experimentally investigated the THz dielectric function of ZnTe under sub-bandgap femtosecond optical excitation at 800 nm wavelength, i.e. under conditions typical for the THz generation in ZnTe. The time- and pump-fluence dependent THz dielectric function, measured in the frequency range 0.5 – 2.5 THz, is very well described by the multi-term Drude-Lorentz model Eq. 1, describing the behavior of the TPA-generated free carriers, and three THz-active phonon modes: two combination modes at $f_{L1} = 1.55$ THz and $f_{L2} = 3.69$ THz, and a fundamental TO-mode at $f_{TO} = 5.32$ THz. We have extracted the free carrier dc conductivity $\sigma_{dc}$, generated free carrier density *N*, and density-dependent electron momentum scattering time $\tau_e(N)$. We have also characterized the dynamics of the THz-active phonon modes following the photoexcitation of ZnTe and have quantified the pump-induced phonon mode activation, which we attribute to the coupling between the lattice and the transient electron gas. Our findings will have direct impact on the accurate, detailed description of the THz generation process in ZnTe by 800 nm excitation, one of the most widely used methods of THz generation nowadays.

**Acknowledgements**

We acknowledge the financial support from the European Union's Horizon 2020 research and innovation program (Grant Agreement No. 964735 EXTREME-IR), Deutsche Forschungsgemeinschaft (DFG) within Project No. 468501411-SPP2314 INTEGRATECH and Project No. 518575758 HIGHSPINTERA, Bundesministerium für Bildung und Forschung (BMBF) within Project No. 05K2022 PBA Tera-EXPOSE, and Bielefelder Nachwuchsfonds.

## Appendix

### Section A1

The optical pump creates a non-uniform excitation profile within the ZnTe crystal due to attenuation along the propagation axis. To quantify this depth dependence, we consider our highest applied pump fluence of 0.5 mJ/cm² (incident peak intensity $I_0 \approx 4.0$ GW/cm²). The effective propagation depth $d_{eff}$, where the pump intensity decreases to $I_0/e$, is given by:

$$d_{eff} = \frac{e-1}{\beta I_0}$$

Using our determined two-photon absorption coefficient $\beta = 4.30$ cm/GW, we find $d_{eff} \approx$ 1 mm. Given the crystal thickness of $d = 0.5$ mm, the pump intensity at the back face drops to approximately 54% of its initial value. Because carrier generation scales quadratically with intensity, the carrier density at the back face drops to roughly 29% of the front-face value. For smaller fluences, the pumping profile becomes substantially more uniform.

Despite this spatial gradient, analytical multi-layer modeling of the THz transfer function indicates that the cascaded propagation and Fresnel transmission terms can be robustly approximated by an effective, homogeneous refractive index. Error propagation confirms that any mathematical deviations introduced by this uniform approximation are negligible compared to the statistical uncertainty of our measurements.

Because the transmitted THz probe pulse is predominantly sensitive to the integrated sheet conductivity, our uniform extraction model inherently performs a spatial average. Therefore, all extracted parameters correspond to macroscopic quantities averaged over the entire crystal thickness $d$:

$$X_{eff} = \frac{1}{d}\int_0^d X\,(z)dz$$

Consequently, the assumption of a uniform effective excitation profile yields valid, thickness-averaged dynamics without introducing significant artifacts.

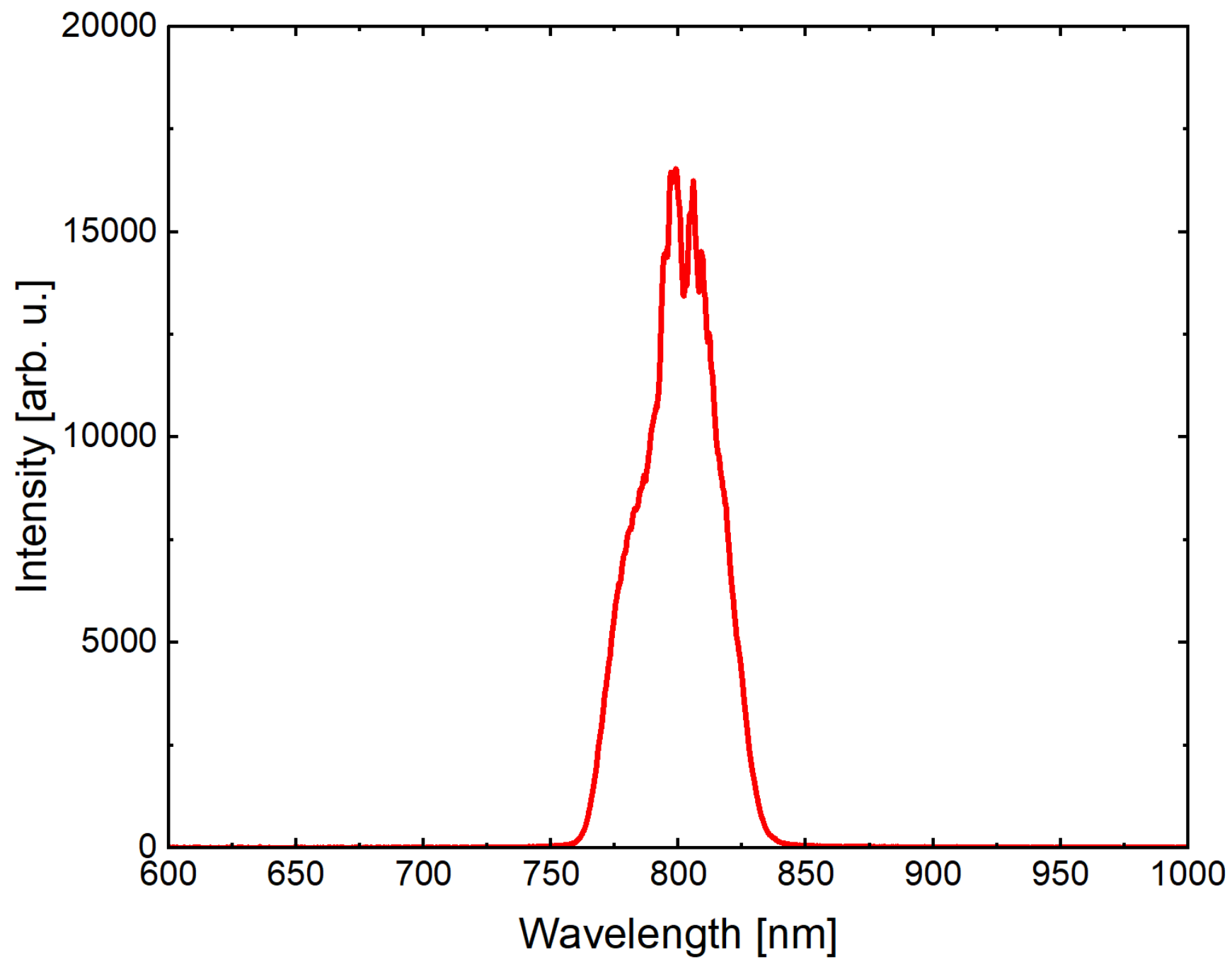

Fig. A1: The spectrum of the Ti:sapphire laser used in our experiments.

**Section A2**

As shown in the main text (Fig. 3(a)), the unconstrained Drude-Lorentz model provides an excellent agreement with the measured transient dielectric function across all pump-probe delays. To ensure maximum transparency regarding the extracted parameters, we analyzed the parameter correlation matrix (Fig. A2(b)). We observe a strong mathematical anti-correlation ($r \approx -0.99$) specifically between the high-frequency dielectric constant ($\varepsilon_\infty$) and the oscillator strength ($\Omega_{L2}$) of the mode L2. This degeneracy arises because, within our limited probing bandwidth, both parameters act as coupled offsets to the real part of the dielectric function.

Typically, to avoid parameter cross-talk, $\varepsilon_\infty$ is fixed to its unpumped reference value. However, applying this constraint to our transient data at positive delays (e.g., at 1 ps) results in a failure of the fitting model, see Fig. A2(a). This indicates that the optical excitation induces a massive, broadband change to the dielectric background that cannot be captured if $\varepsilon_\infty$ is kept static.

Therefore, both $\varepsilon_\infty$ and $\Omega_{L2}$ were kept as free fitting parameters for all delay times. Consequently, their individual transient dynamics, shown in Fig. A2(c) and Fig. A3(b), exhibit strong mirroring effects due to the $r \approx -0.99$ correlation. We emphasize that these two parameters should not be interpreted as isolated physical quantities, but rather as an effective, coupled phenomenological response that correctly scales the baseline and amplitude of the resonance upon optical pumping. Crucially, this correlation does not affect the precision of the uncorrelated parameters discussed in the main text, such as the oscillator strength and linewidth of the phonon mode at $f_{L1} = 1.55$ THz or the dc conductivity.

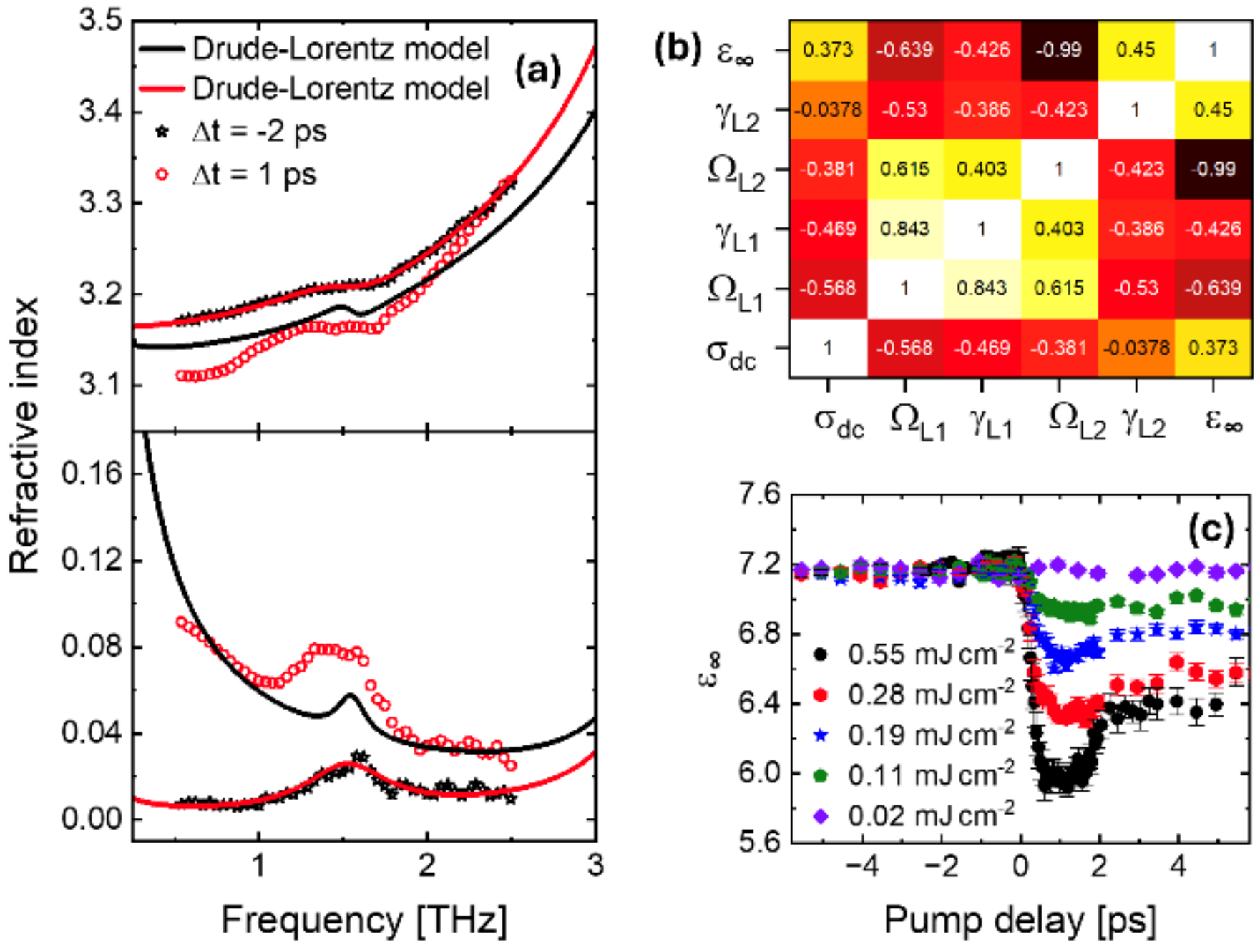

Fig. A2: The measured complex refractive index 2 ps before and 1.3 ps after the excitation, at the pump fluence of 0.55 mJ cm$^{-2}$. The refractive index is depicted in the top, the extinction coefficient in the bottom part of the graph. The experimental data is displayed as symbols and solid lines represent the fitting results when $\varepsilon_\infty$ is fixed. (b) The correlation matrix for the fitting to the dielectric function 2 ps before the excitation. (c) Extracted dynamics of the high-frequency dielectric constant for different pump fluences.

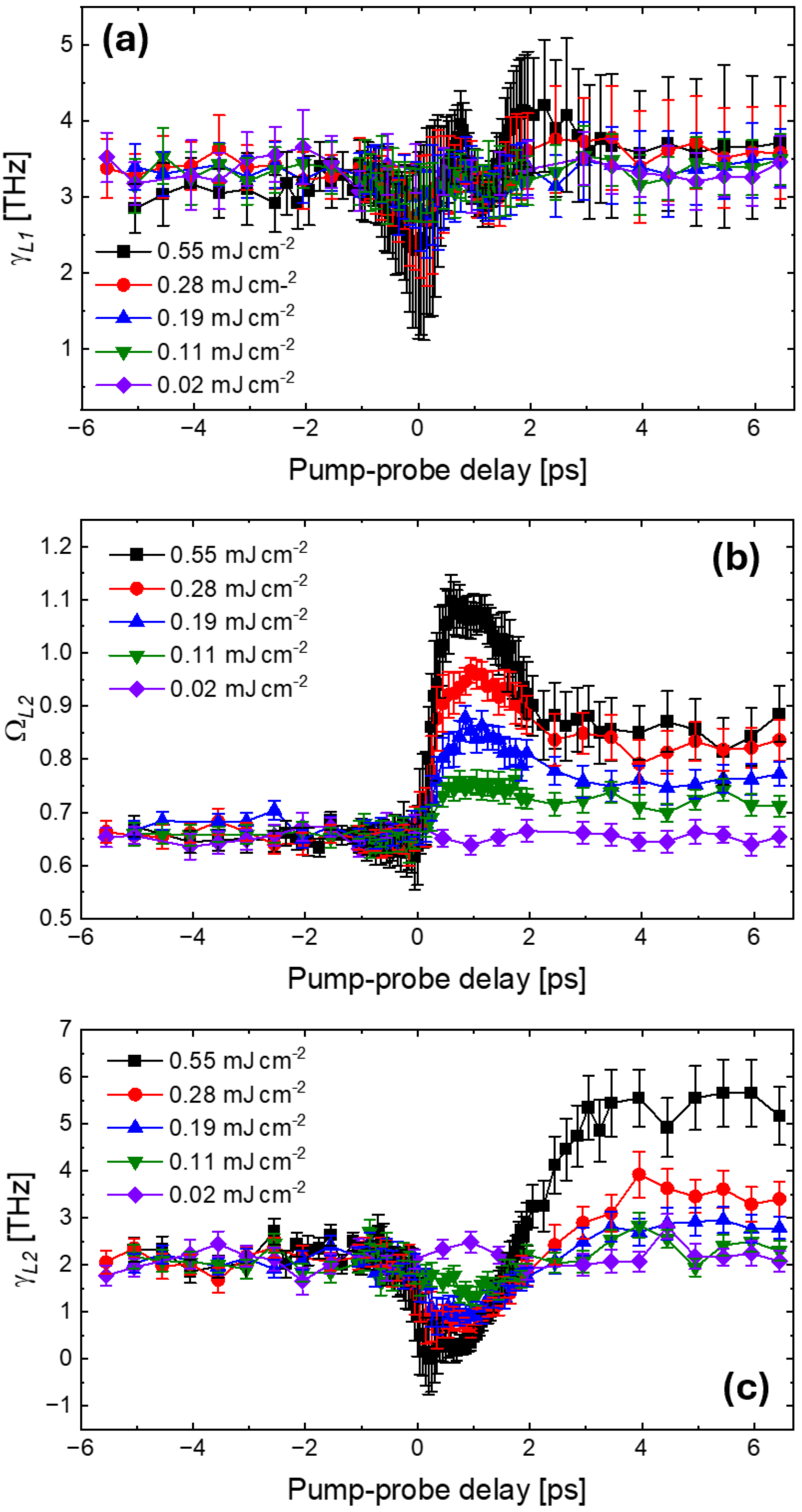

Fig. A3: (a) The extracted linewidth of the L1 phonon mode at $f_{L1} = 1.55$ THz. (b) The extracted oscillator-strength of the L2 phonon mode at $f_{L2} = 3.69$ THz. (c) The linewidth of the oscillator-strength of the L2 phonon mode at $f_{L2} = 3.69$ THz.

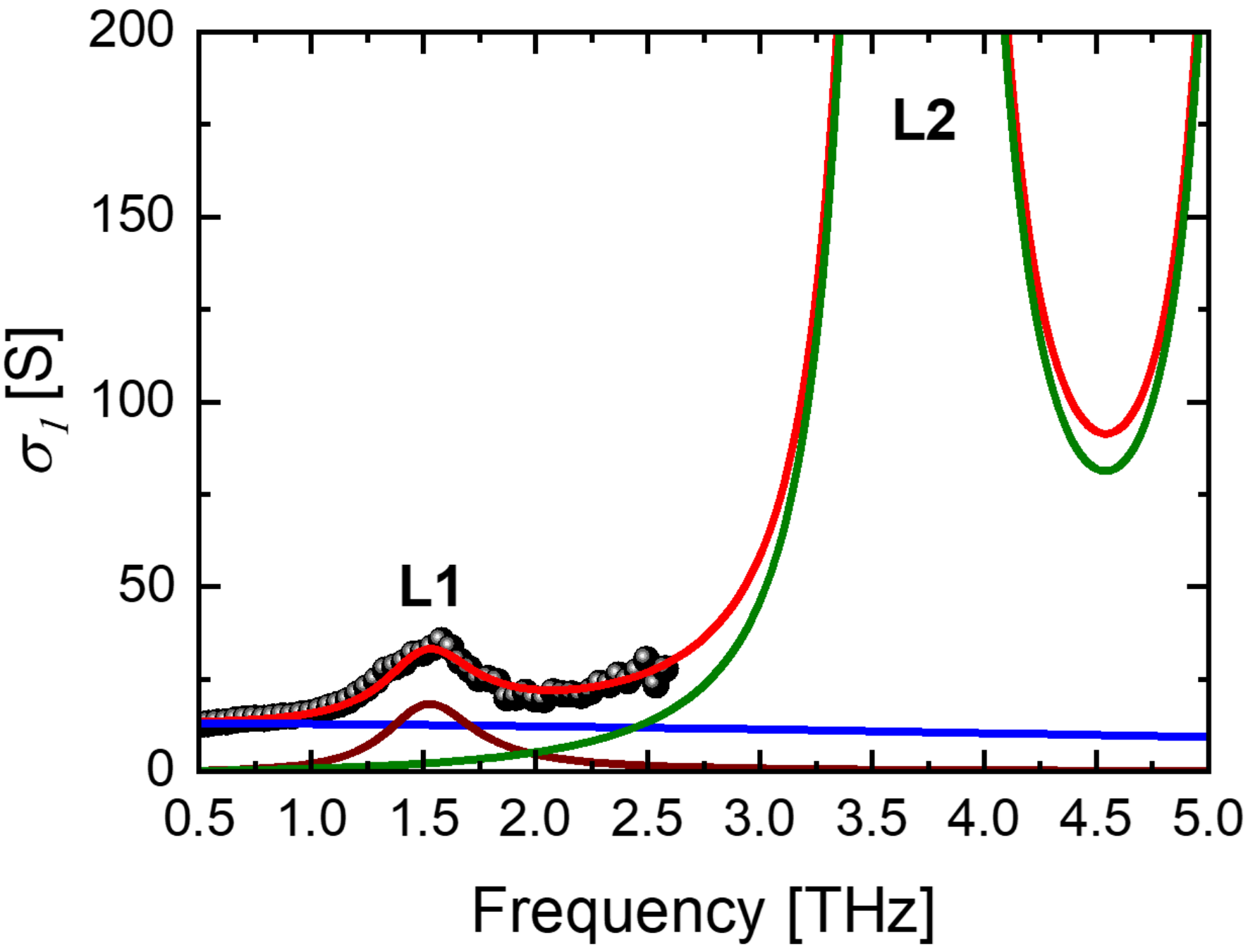

Fig. A4: The partial fits in the real part of the conductivity up to 5 THz. The black symbols represent the measured data $\Delta t =$ 1.3 ps after the excitation with the pump fluence of 0.55 mJ $cm^{-2}$, the red solid line represents the total fit, the blue line the Drude contribution, the reddish line the contribution by resonance L1 and the green line the background conductivity by resonance L2 and the TO-phonon.